\documentclass[manuscript]{aastex}
\usepackage{natbib}
\usepackage{booktabs}
\usepackage{threeparttable}
\usepackage{amsmath,bm}
\begin{document}

\title{Dependence of ion temperatures on alpha$-$proton differential flow vector and heating mechanisms in the solar wind}

\author{G. Q. Zhao\altaffilmark{1,2,3}, H. Q. Feng\altaffilmark{1}, D. J. Wu\altaffilmark{4}, J. Huang\altaffilmark{2}, Y. Zhao\altaffilmark{1}, Q. Liu\altaffilmark{1}, and Z. J. Tian\altaffilmark{1}}

\affil{$^1$Institute of Space Physics, Luoyang Normal University, Luoyang, China}
\affil{$^2$CAS Key Laboratory of Solar Activity, National Astronomical Observatories, Beijing, China}
\affil{$^3$Henan Key Laboratory of Electromagnetic Transformation and Detection, Luoyang, China}
\affil{$^4$Purple Mountain Observatory, CAS, Nanjing, China}

\begin{abstract}
According to \emph{Wind} observations between June 2004 and May 2019, this Letter investigates the proton and alpha particle temperatures in the space of ($\theta_d$, $V_d/V_A$) for the first time, where $\theta_d$ and $V_d$ are the radial angle and magnitude of alpha$-$proton differential flow vector ${\bm V}_d$ respectively, $V_A$ is the local Alfv\'en speed. Results show that the temperatures significantly depend on $\theta_d$ as well as $V_d/V_A$. In case of low proton parallel beta ($\beta_{p{\parallel}} < 1$), it is found that the proton perpendicular temperature is clearly enhanced when $\theta_d$ is small ($\lesssim 45^\circ$) and $V_d/V_A \gtrsim 0.5$. On the contrary, the perpendicular temperature of alpha particles is considerably enhanced when $\theta_d$ is large ($\gtrsim 90^\circ$) or $V_d/V_A$ is sufficiently small. The maximum of proton parallel temperature takes place at $\theta_d \sim 90^\circ$ accompanied by higher $\beta_{p{\parallel}}$ and by larger turbulence amplitude of magnetic fluctuations in inertial range. This study should present strong evidence for cyclotron resonance heating of protons and alpha particles in the solar wind. Other mechanisms including Landau resonance and stochastic heating are also proposed, which tend to have different ($\theta_d$, $V_d/V_A$) spaces than cyclotron resonance heating.
\end{abstract}


\section{Introduction}
The solar wind is a tenuous, magnetized plasma streaming outward from the Sun \citep[e.g.,][]{han12p89,abb16p55}. It is highly nonadiabatic with ion temperatures higher than those from a spherically expanding ideal gas, implying that some heating process must occur in the solar wind \citep[e.g.,][]{hun70p43,bam75p73,fel98p47,sta19p02}.
Although the process is poorly understood, it is fundamentally important to describe the solar wind and to characterize astrophysical plasmas more generally. Many mechanisms have been proposed in terms of wave-particle resonances \citep[cyclotron resonance and Landau resonance;][]{mar82p30,hol02p47,cra14p16,hej15p76,hej15p31,how18p05}, or non-resonant stochastic heating \citep[][]{joh01p21,wan06p01,wuc07p01,yoo09p02,bou13p96,mar19p43}, or plasma coherent structures such as magnetic vortices, reconnecting current sheets, and shocks \citep{bru03p30,osm12p02,per12p01,wan19p22}. Theoretically, these mechanisms are characterized by different properties. The cyclotron resonance mechanism, for instance, will produce perpendicular heating of ions with respect to the background magnetic field. The stochastic heating mechanism will also lead to perpendicular heating of ions, which requires a large turbulence amplitude satisfying some critical value \citep[e.g.,][]{cha10p03,vec17p11}. The Landau resonance mechanism, on the other hand, can contribute to parallel heating of protons.

The solar wind is generally far from thermodynamic equilibrium \citep[e.g.,][]{mar82p35,mar82p52,alt18p12,zha19p75}. Alpha particle populations usually have different bulk velocities with respect to protons, and hence a differential flow arises. The differential flow vector is defined as ${\bm V}_d = {\bm V}_\alpha -{\bm V}_p$ throughout this Letter, where ${\bm V}_\alpha$ and ${\bm V}_p$ are proton and alpha particle bulk velocities, respectively. The ${\bm V}_d$ points anti-Sunward when its radial angle satisfies $0^\circ \leq \theta_d < 90^\circ$, while it is directed Sunward if $90^\circ < \theta_d \leq 180^\circ$ is fulfilled. In literatures the alpha$-$proton differential flow was intensively discussed in terms of its magnitude ($V_d$). Theory and simulations revealed that a large $V_d$ normalized by local Alfv\'en speed $V_A$ can excite kinetic waves that then heat the solar wind \citep{gar00p20,luq06p01,gao13p71}. In situ measurements further showed that $V_d/V_A$ can regulate the (relative) temperatures of protons and alpha particles in the solar wind \citep{kas08p03,kas13p02}. \citet[][their Figure 5]{kas08p03}, for example, particularly showed that $T_{\alpha\perp}/T_{p\perp}$ decreases as $V_d/V_A$ increases for the solar wind with infrequent collisions. This result is well in line with the theory that an increasing $V_d/V_A$ will weaken the alpha cyclotron resonance \citep{ise83p23,gar01p55}.

Less attention, however, has been paid to the flow direction with respect to the Sun. The flow ${\bm V}_d$ is a vector that often points anti-Sunward but sometimes is directed Sunward \citep{fuh18p84,zha19p00}. The flow direction is perhaps inherently important. Previous studies revealed that the flow direction with respect to the propagation direction of Alfv\'en-cyclotron fluctuations is a critical factor in determining the wave-particle interactions in terms of cyclotron resonance. Through investigating the cyclotron resonance factors and the associated damping of fluctuations, \citet{gar05p08,gar06p05} demonstrated that the alpha cyclotron resonance is strong when the flow direction is opposite to the propagation direction of the fluctuations. If they have the same direction, the alpha cyclotron resonance is strong only when $V_d/V_A$ is sufficiently small. Specifically, the alpha cyclotron resonance gradually weakens as $V_d/V_A$ increases from zero to 0.5, above which the proton cyclotron resonance almost completely dominates.

In this Letter, based on in situ measurements, we report our finding that ion (proton and alpha particle) temperatures show significant dependence on the direction of ${\bm V}_d$. In particular, it is shown that proton perpendicular temperature is distinctly large when ${\bm V}_d$ points prominently anti-Sunward with $V_d/V_A \gtrsim 0.5$, while the perpendicular temperature of alpha particles is enhanced when ${\bm V}_d$ is directed Sunward or $V_d/V_A$ is small. This may provide crucial indication for cyclotron resonance heating of ions in the solar wind. Meanwhile it also tends to suggest that the ($\theta_d$, $V_d/V_A$) space is a helpful space to discuss other heating mechanisms including Landau resonance and stochastic heating in the solar wind, where $\theta_d$ is the angle between ${\bm V}_d$ and the solar wind bulk velocity whose direction is represented by the radial vector of the Sun in this Letter. The observations and data analysis are described in Section 2. A summary with brief discussion is presented in Section 3.

\section{Observations and data analysis}
The plasma data used in this Letter are from the Solar Wind Experiment (SWE) instrument on board the \emph{Wind} mission \citep{ogi95p55}. They are produced via a nonlinear-least-squares bi-Maxwellian fit of ion spectrum measured by the Faraday cup with a cadence of 92 s \citep{kas06p05}. This data includes the proton and alpha particle bulk velocities and their perpendicular and parallel temperatures with respect to the background magnetic field ${\bm B}_0$.

The data are chosen between June 2004 and May 2019, during which the \emph{Wind} mission has a halo orbit around the L1 Lagrange point. Other operations are performed to select the data as follows. Firstly, all observations with ${V}_d/{V}_p < 1\%$, which constitute $\sim$ 19\% of the total data set, are discarded because in the case ${V}_d$ would have a large uncertainty \citep{kas06p05,alt18p12}. Secondly, it is required that the angle between ${\bm V}_d$ and ${\bm B}_0$ (or $-{\bm B}_0$) is less than $20^\circ$ since the differential flow is believed to be aligned with ${\bm B}_0$ \citep[e.g.,][]{alt18p12}; this operation leads to a discarding of additional $\sim$ 38\% of the total data set. In addition, the Coulomb collisional age $A_c$ is calculated, which is the ratio of the transit time of the solar wind to the collision timescale \citep{liv86p45}. Observations with $A_c < 0.1$ are selected to obtain the solar wind with a negligible collision effect; this operation additionally excludes $\sim$ 25\% of the total data set. Finally, the sample for the analysis consists of $\sim4.7 \times 10^5$ data, among which $\sim4.6 \times 10^4$ data are available to investigate the alpha particle temperatures.

The proton parallel beta $\beta_{p{\parallel}}$ is an important parameter in theoretical studies of the heating mechanisms, which is defined as the ratio of proton parallel pressure to magnetic pressure. Figure 1 is presented for the case of low beta with $\beta_{p{\parallel}} < 1$, where the data distributions and the medians of perpendicular temperatures for protons and alpha particles are plotted in ($\theta_d$, $V_d/V_A$) space, respectively. Each cell with size $5^\circ \times 0.05$ for protons (left panels) and $10^\circ \times 0.1$ for alpha particles (right panels) is set in the space; here any cell with less than 10 data points is not considered (marked by the white color). From panel (a), one can see that there are a considerable number of observations with the differential flow directed toward the Sun though most observations are with the flow outward. From panel (b), it is clear that the proton perpendicular temperature $T_{p\perp}$ first depends on $V_d/V_A$. Overall, there is a tendency that $T_{p\perp}$ increases with $V_d/V_A$. $T_{p\perp}$ is commonly low (typically $\sim 1.0 \times 10^5 ~K$) when $V_d/V_A$ is small ($< 0.4$) while it can be very high (up to $2.5 \times 10^5 ~K$) when $V_d/V_A > 0.7$; it rises rapidly at $V_d/V_A \sim 0.5$ denoted by the gray dotted line in the figure. On the other hand, $T_{p\perp}$ also depends on the radial angle $\theta_d$, and the highest $T_{p\perp}$ occurs at $\theta_d \sim 0^\circ$. A dependence of alpha perpendicular temperature $T_{\alpha\perp}$ (normalized by $T_{p\perp}$ as usual) on $V_d/V_A$ and $\theta_d$ also appears. Panel (d) of Figure 1 shows that a smaller $V_d/V_A$ and a larger $\theta_d$ will correspond to a higher $T_{\alpha\perp}/T_{p\perp}$; $T_{\alpha\perp}/T_{p\perp}$ has its highest values up to 7 at $V_d/V_A < 0.2$ and $\theta_d > 150^\circ$.

\begin{figure}
\epsscale{0.9} \plotone{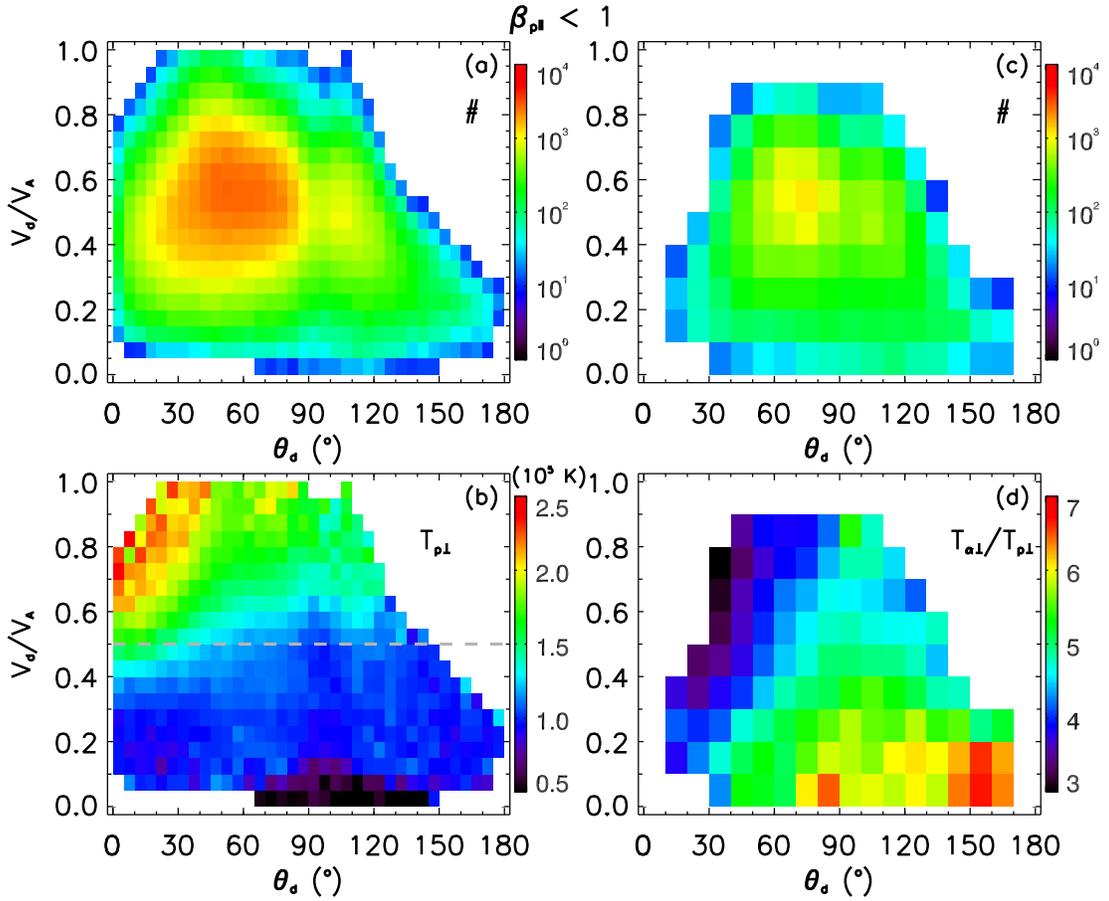} \caption{Data distributions and medians of ion perpendicular temperatures with $\beta_{p{\parallel}} < 1$: panel (a), sample number distribution for proton temperature;  panel (b), proton perpendicular temperature $T_{p\perp}$; panel (c), sample number distribution for alpha particle temperature; panel (d), alpha perpendicular temperature relative to that of protons $T_{\alpha\perp}/T_{p\perp}$.}
\end{figure}

\begin{figure}
\epsscale{0.9} \plotone{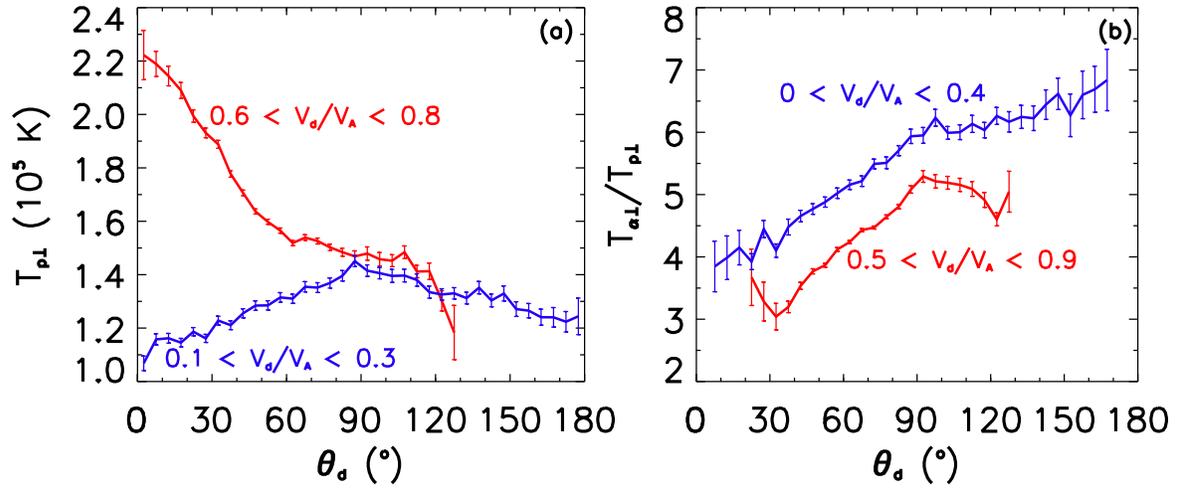} \caption{$T_{p\perp}$  and $T_{\alpha\perp}/T_{p\perp}$ with respect to the radial angle $\theta_d$, where the red line represents the temperature with large $V_d/V_A$ while the blue line refers to the temperature with small $V_d/V_A$ in each panel.}
\end{figure}

To further illustrate the dependence of the temperatures on the flow direction revealed in Figure 1, Figure 2 displays mean values of $T_{p\perp}$ (panel (a)) and $T_{\alpha\perp}/T_{p\perp}$ (panel (b)) against $\theta_d$ for different ranges of $V_d/V_A$. In each panel the red line represents the temperature with large $V_d/V_A$ while the blue line refers to the temperature with small $V_d/V_A$. Here any $T_{\alpha\perp}/T_{p\perp}$ with an uncertainty larger than $0.5$ is not shown; to produce the low uncertainty wider ranges of $V_d/V_A$ are used in panel (b). One can see that $T_{p\perp}$ bounded by $0.6 < V_d/V_A < 0.8$ (the red line in panel (a)) and $T_{\alpha\perp}/T_{p\perp}$ bounded by $0 < V_d/V_A < 0.4$ (the blue line in panel (b)) show clear but different dependences on $\theta_d$. The $T_{p\perp}$ decreases from ($2.22 \pm 0.09$) $\times 10^5 ~K$ to ($1.47 \pm 0.02$) $\times 10^5 ~K$ as $\theta_d$ increases from 0$^\circ$ to approaching 90$^\circ$, while the $T_{\alpha\perp}/T_{p\perp}$ almost monotonously increases from 3.85 $\pm$ 0.41 to 6.84 $\pm$ 0.49 with $\theta_d$ from 0$^\circ$ to exceeding 160$^\circ$.

Figure 3 presents the case of $\beta_{p{\parallel}} > 1$ with the same format as Figure 1. Comparing Figure 3 with Figure 1, a similar result is that $T_{p\perp}$ is higher when $V_d/V_A \gtrsim 0.5$ than that when $V_d/V_A < 0.5$ and a higher $T_{\alpha\perp}/T_{p\perp}$ results from a smaller $V_d/V_A$ and a larger $\theta_d$ in principle. Here it is also interesting that $T_{p\perp}$ in the region of $\theta_d \sim 90^\circ$ is comparable to that with a small $\theta_d$, which is very different from the case in Figure 1 since $T_{p\perp}$ in Figure 1 rapidly decreases with $\theta_d$ when $V_d/V_A \gtrsim 0.5$. Moreover, $T_{\alpha\perp}/T_{p\perp}$ in Figure 3 seems to be also considerably enhanced when $\theta_d \sim 90^\circ$ for a very large $V_d/V_A$ ($\sim 1$). The enhancement may be attributed to non-resonant stochastic heating that is believed to work efficiently with a large turbulence amplitude. This will be discussed later.

\begin{figure}
\epsscale{0.9} \plotone{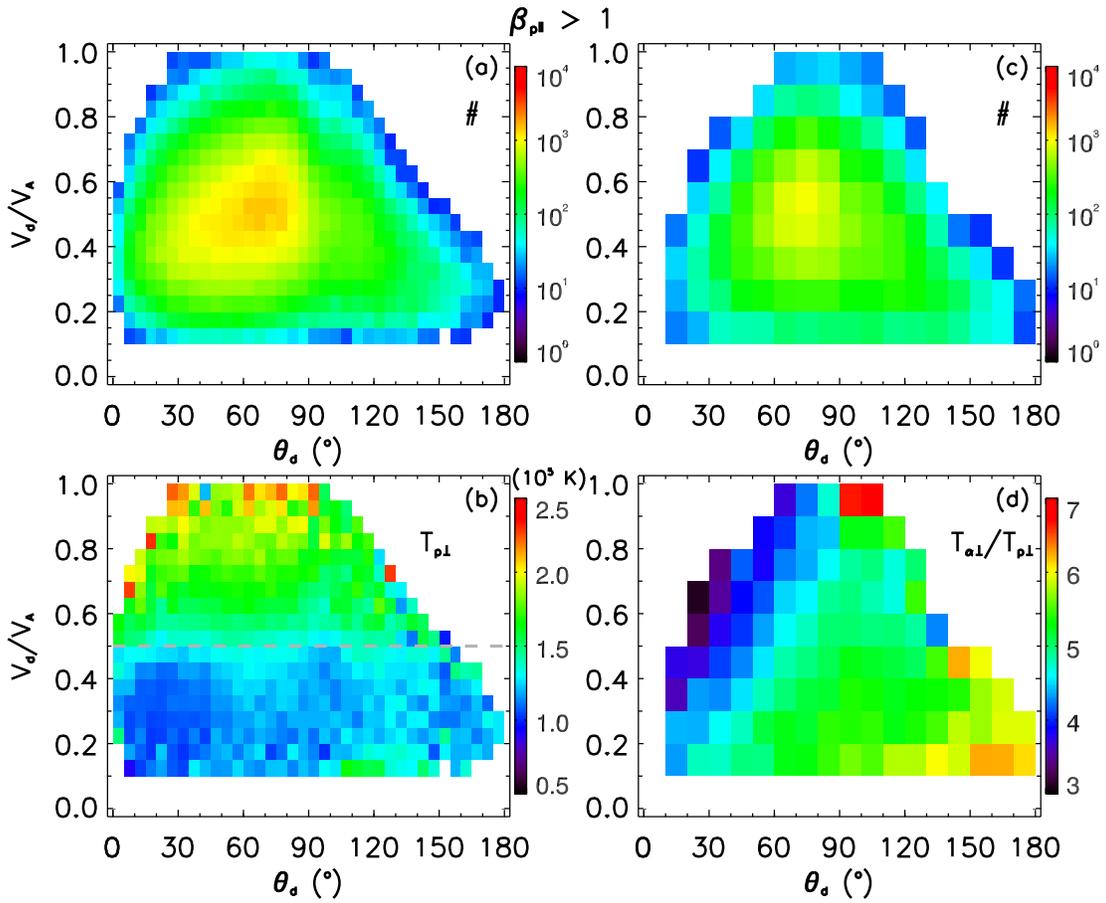} \caption{Data distributions and medians of ion perpendicular temperatures with $\beta_{p{\parallel}} > 1$: panel (a), sample number distribution for proton temperature;  panel (b), proton perpendicular temperature $T_{p\perp}$; panel (c), sample number distribution for alpha particle temperature; panel (d), alpha perpendicular temperature relative to that of protons $T_{\alpha\perp}/T_{p\perp}$.}
\end{figure}

A dependence on the differential flow also appears for proton parallel temperature $T_{p{\parallel}}$. Figure 4 plots medians of $T_{p{\parallel}}$ sorted by ($\theta_d$, $V_d/V_A$) for $\beta_{p{\parallel}} < 1$ (panel (a)) and $\beta_{p{\parallel}} > 1$ (panel (d)). One can see that $T_{p{\parallel}}$ depends on both $V_d/V_A$ and $\theta_d$. Different from the result for $T_{p\perp}$, higher $T_{p{\parallel}}$ occurs particularly at $\theta_d \sim 90^\circ$, and this situation is regardless of $\beta_{p{\parallel}} < 1$ or $> 1$. (This kind of dependence is not clear for alpha particles (not shown).) Moreover, the increase of $T_{p{\perp}}$ is rapid at $V_d/V_A \sim 0.5$, while it is mild for $T_{p{\parallel}}$ at the same $V_d/V_A $.

Theoretically, an increase of proton parallel temperature can be due to Landau resonance of protons with kinetic waves or turbulence, which invokes a high $\beta_{p{\parallel}}$ ($\gtrsim 1$). To test the idea, Figure 4 also plots medians of $\beta_{p{\parallel}}$ and turbulence amplitude $\delta{B}$ in the ($\theta_d$, $V_d/V_A$) space, where left panels are for $\beta_{p{\parallel}} < 1$ while right panels are for $\beta_{p{\parallel}} > 1$. The magnetic field data with a cadence of 0.092 s are used, which are from the Magnetic Field Investigation (MFI) instrument on board the \emph{Wind} mission \citep{lep95p07}. Here $\delta{B}$ refers to the average amplitude of turbulence spectrum of magnetic fluctuations in the range of 0.01$-$0.1 Hz (inertial range). To obtain $\delta{B}$, the power spectral density of each magnetic field component is calculated via Fourier transform. This method was used first by \citet{vec18p04}, who proposed the $\delta{B}$ is simple and effective proxy to discuss the connection between the large-scale dynamics of turbulence cascade and particle heating at kinetic scales.

Based on Figure 4, one may speculate that a high $\beta_{p{\parallel}}$ and meanwhile a large $\delta{B}$ are two necessary conditions to generate strong proton parallel heating. First of all, $T_{p{\parallel}}$ with $\beta_{p{\parallel}} > 1$ (panel (d)) is considerably larger than that with $\beta_{p{\parallel}} < 1$ (panel (a)), and the former corresponds to larger $\delta{B}$ (panel (f)) relative to that in panel (c). In case of $\beta_{p{\parallel}} < 1$, a region of ($\theta_d$, $V_d/V_A$) is marked by the gray dotted box in panels (a), (b) and (c) to highlight the speculation. $T_{p{\parallel}}$ is enhanced in the box, which is accompanied by $\beta_{p{\parallel}} > 0.5$ and $\delta{B} \gtrsim 1$ nT$^2$/Hz. Out of the box, on the other hand, either just high $\beta_{p{\parallel}}$ or large $\delta{B}$ tends to fail to produce a high $T_{p{\parallel}}$. The $\beta_{p{\parallel}}$, for instance, is high ($\gtrsim 0.7$) in the region ($\theta_d$, $V_d/V_A$) $\sim$ (150$^\circ$, 0.3), indicated by the gray cross sign in the figure, but $\delta{B}$ is low ($\lesssim$ 0.5 nT$^2$/Hz) relative to that in the box. Consequently, a low $T_{p{\parallel}}$ ($\sim 1 \times 10^5 ~K$) remains.

\begin{figure}
\epsscale{0.9} \plotone{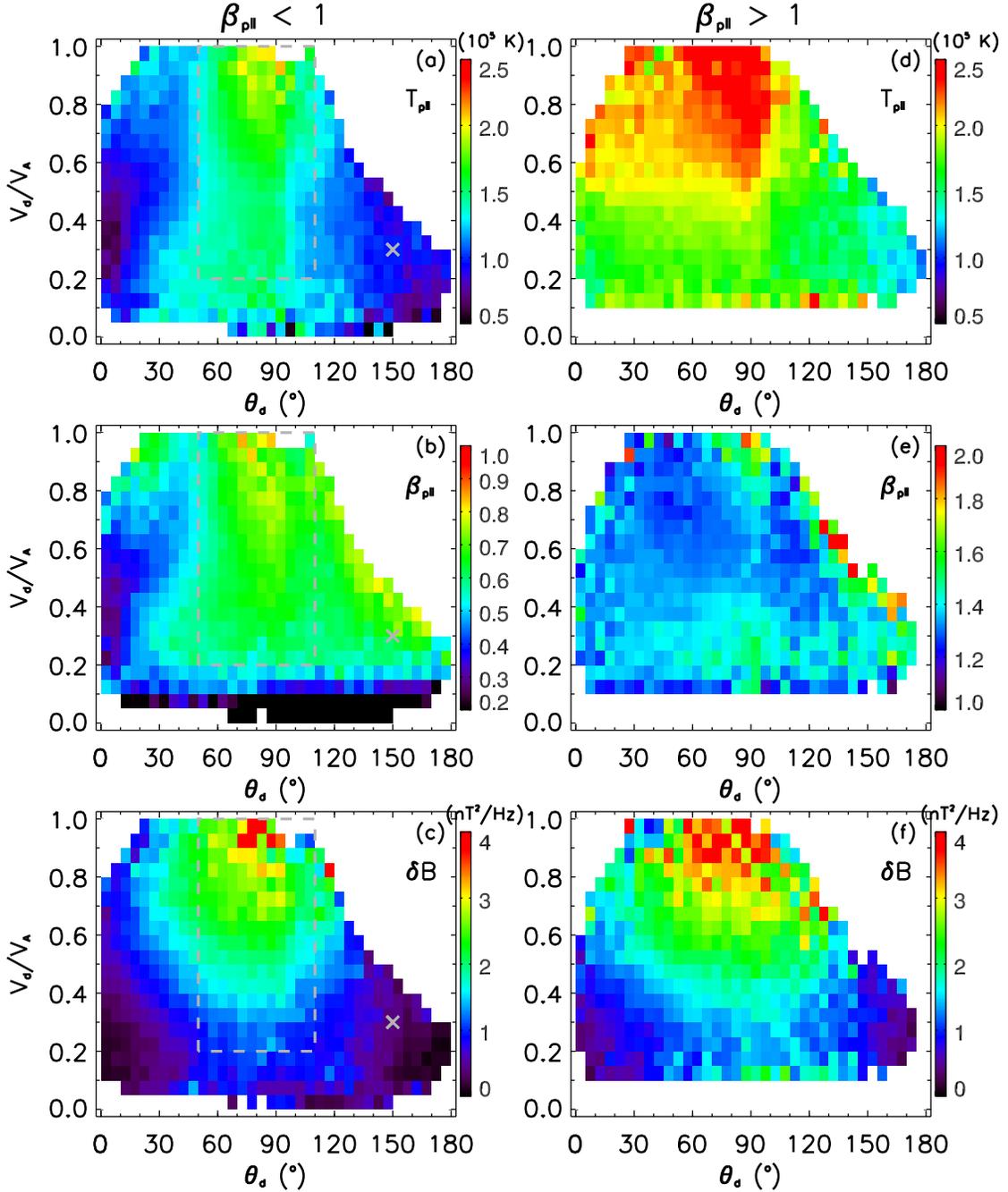} \caption{Medians of proton parallel temperature $T_{p\parallel}$ (top panels), proton parallel beta $\beta_{p\parallel}$ (middle panels), and turbulence amplitude of magnetic fluctuations $\delta{B}$ (bottom panels). Left panels are for $\beta_{p{\parallel}} < 1$ while right panels are for $\beta_{p{\parallel}} > 1$. In left panels the gray dotted box and the cross sign indicate two regions with $T_{p\parallel}$ enhanced or not, respectively.}
\end{figure}

\begin{figure}
\epsscale{0.9} \plotone{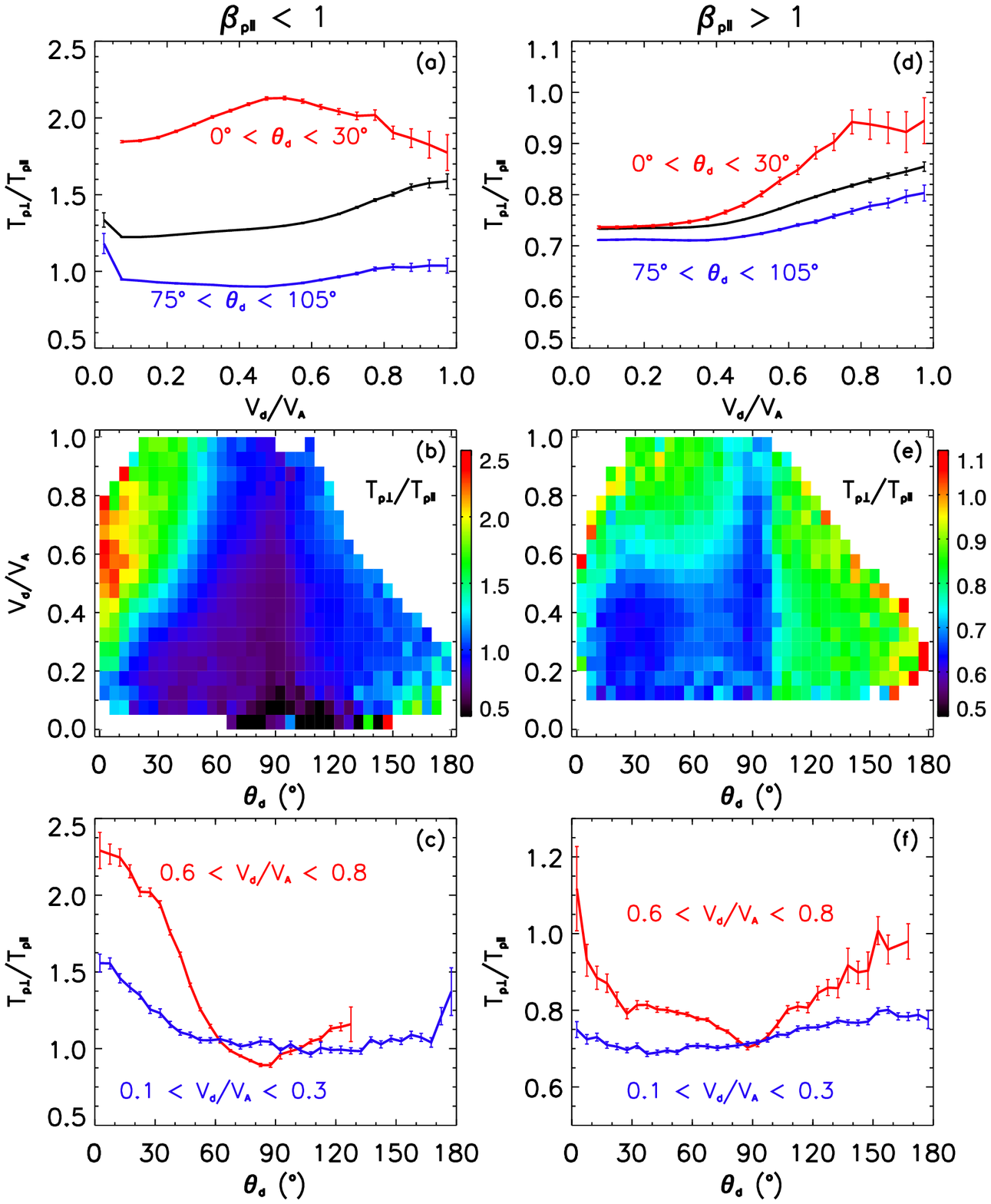} \caption{Mean values of proton temperature anisotropy $T_{p\perp}/T_{p\parallel}$ with respect to $V_d/V_A$ (top panels), medians of proton temperature anisotropy $T_{p\perp}/T_{p\parallel}$ (middle panels), and mean values of proton temperature anisotropy $T_{p\perp}/T_{p\parallel}$ with respect to $\theta_d$ (bottom panels). Left panels are for $\beta_{p{\parallel}} < 1$ while right panels are for $\beta_{p{\parallel}} > 1$.}
\end{figure}

Figure 5 displays the proton temperature anisotropy ($T_{p\perp}/T_{p{\parallel}}$) sorted by $V_d/V_A$ and/or $\theta_d$, where left panels are for $\beta_{p{\parallel}} < 1$ and right panels are for $\beta_{p{\parallel}} > 1$. Previous studies revealed that $T_{p\perp}/T_{p{\parallel}}$ slightly increases in principle with $V_d/V_A$ for the solar wind \citep[e.g.,][]{kas08p03}. The present data are in line with this result as shown by the black lines in panels (a) and (d), where mean values of $T_{p\perp}/T_{p{\parallel}}$ are plotted against $V_d/V_A$, black lines are for $T_{p\perp}/T_{p{\parallel}}$ without limit of $\theta_d$, red lines are for $T_{p\perp}/T_{p{\parallel}}$ bounded by $0^\circ < \theta_d < 30^\circ$, blue lines are for $T_{p\perp}/T_{p{\parallel}}$ requiring $75^\circ < \theta_d < 105^\circ$. Comparing the red lines with the blue lines, on the other hand, one can find that $T_{p\perp}/T_{p{\parallel}}$ with a small $\theta_d$ is generally larger than that with a medium $\theta_d$. In addition, nonmonotonic $T_{p\perp}/T_{p{\parallel}}$ with respect to $V_d/V_A$ appears in case of $\beta_{p{\parallel}} < 1$ and $0^\circ < \theta_d < 30^\circ$, as shown by the red line in panel (a). (Note that nonmonotonic $T_{p\perp}/T_{p{\parallel}}$, initially increasing and then decreasing as $V_d/V_A$ varying from 0 to 0.75, was obtained by \citet{gar06p05} with a topic of Alfv\'en-cyclotron scattering of ions in terms of hybrid simulations for a low-beta solar wind, their Figure 7(a).)
More complete picture of $T_{p\perp}/T_{p{\parallel}}$ (medians) against $\theta_d$ and $V_d/V_A$ is displayed by panels (b) and (e), showing that $T_{p\perp}/T_{p{\parallel}}$ depends on $\theta_d$ as well as $V_d/V_A$. The maximum of $T_{p\perp}/T_{p{\parallel}}$, up to 2.42, occurs at ($\theta_d$, $V_d/V_A$) $\sim$ (0$^\circ$, 0.6) while its minimum, low to 0.33, tends to take place when $\theta_d$ is medium ($\sim 90^\circ$) and $V_d/V_A$ is very small ($\sim 0.05$) in case of $\beta_{p{\parallel}} < 1$ (panel (b)). In case of $\beta_{p{\parallel}} > 1$, $T_{p\perp}/T_{p{\parallel}}$ is usually less than unity with a relatively weak dispersion, as shown in panel (e). Panels (c) and (f) plot the mean values of $T_{p\perp}/T_{p{\parallel}}$ with respect to $\theta_d$ for reference, where red lines are with $0.6 < V_d/V_A < 0.8$ while blue lines require $0.1 < V_d/V_A < 0.3$. A result is that a comparable $T_{p\perp}/T_{p{\parallel}}$ appears when $\theta_d$ approaches $90^\circ$ especially for the case $\beta_{p{\parallel}} > 1$ (panel (f)).

\section{Summary and discussion}
Based on in situ measurements lasting for 15 years in the solar wind, this Letter carries out a statistical study on proton and alpha particle temperatures in ($\theta_d$, $V_d/V_A$) space for the first time. Results show that the temperatures depend on not only the amplitude but also the direction of alpha$-$proton differential flow. The proton perpendicular temperature is clearly enhanced when the flow is directed sufficiently anti-Sunward ($\theta_d \lesssim 45^\circ$) and the amplitude is large ($V_d/V_A \gtrsim 0.5$) in the low-beta solar wind. On the contrary, the alpha perpendicular temperature is enhanced preferentially with Sunward direction or sufficiently small amplitude of the flow. In addition, the proton parallel temperature has its maximum when the flow is tangential ($\theta_d \sim 90^\circ$). Further investigation on the proton parallel beta and turbulence amplitude of magnetic fluctuations shows that the rise of the parallel temperature corresponds to the higher beta and larger turbulence amplitude. The proton temperature anisotropy also depends on the amplitude and direction of alpha$-$proton differential flow. In case of low beta, a region ($\theta_d$, $V_d/V_A$) $\sim$ (0$^\circ$, 0.6) corresponds to a strong temperature anisotropy (with median up to 2.42), and another region ($\theta_d$, $V_d/V_A$) $\sim$ (90$^\circ$, 0.05) corresponds to a converse strong temperature anisotropy (with median low to 0.33).

The dependence revealed in Figures (1) and (2) may be readily understood in terms of ion cyclotron resonance heating. The cyclotron resonance heating is believed to first increase the perpendicular temperature of ions. Based on linear Vlasov theory and hybrid simulation, \citet{gar05p08,gar06p05} have demonstrated that the differential flow amplitude and its direction with respect to the propagation direction of Alfv\'en-cyclotron fluctuations sensitively determine the ion cyclotron resonance properties and therefore the cyclotron damping of such fluctuations. If they have the same direction, alpha cyclotron resonance is significant only when the flow amplitude is sufficiently small ($V_d/V_A \ll 1$), and proton cyclotron resonance will become significant when the flow amplitude is large. In particular, it was shown that the damping due to proton cyclotron resonance almost completely dominates once $V_d/V_A \gtrsim 0.5$ is fulfilled. If their directions are opposite, alpha cyclotron resonance dominates, and consequently the heating of alpha particles should be expected. Here, one may speculate that the related Alfv\'en-cyclotron fluctuations are mainly outward propagations. This should be reasonable because the Alfv\'en waves in the solar wind have been shown as mainly outward propagations \citep{hej09p25,lih1602,yan17p77}. These Alfv\'en waves probably come from the Sun, and as they propagate outward they will evolve to be Alfv\'en-cyclotron waves via a frequency sweeping process due to the gradually reduced background magnetic field. Moreover, \citet{rob15p01} demonstrated that Alfv\'en-cyclotron waves in their wave sample are outward based on dispersion relation analysis with $k$-filtering technique. According to the researches by \citet{gar05p08,gar06p05}, the outward Alfv\'en-cyclotron waves could significantly heat protons when the differential flows is directed anti-Sunward with a large amplitude and mainly heat alpha particles when the differential flow is directed Sunward, as shown in Figures (1) and (2). In this regard, the present study offers strong evidence for cyclotron resonance heating of ions based on direct temperature measurements. 

In addition, proton Landau resonance heating also likely works in the solar wind according to the present study (Figure 4). To be precise, this process tends to preferentially occur at a region where the alpha$-$proton differential flow is quasi-perpendicular to the radial vector of the Sun. In the region the proton parallel beta is higher and simultaneously the turbulence amplitude is larger.

Moreover, this study may present hint for non-resonant stochastic heating of ions, which is believed to increase the perpendicular temperatures of protons and alpha particles depending on the turbulence amplitude (ion gyroscale exactly) \citep{cha10p03,hop18p15}. The rises of perpendicular temperatures of protons and alpha particles at the region ($\theta_d$, $V_d/V_A$) $\sim$ ($90^\circ$, 0.9), which is weak in Figure 1 but is strong in Figure 3, could be attributed to the stochastic heating, because at the same region the largest turbulence amplitude of gyroscale fluctuations can be expected according to Figure 4 (bottom panels).

\acknowledgments
The authors thank the SWE team and MFI team on \emph{Wind} for providing the data, which are available via the Coordinated Data Analysis Web (http://cdaweb.gsfc.nasa.gov/cdaweb/istp$_-$public/). This research was supported by NSFC under grant Nos. 41874204, 41974197, 41674170, 41531071, 11873018. Research by G. Q. Zhao was also supported by the Project for Scientific Innovation Talent in Universities of Henan Province (19HASTIT020), and partly by the Key Laboratory of Solar Activity at CAS NAO (KLSA201703). Research by Q. Liu was supported by Key Scientific Research Project in Universities of Henan Province (16B140003). The authors acknowledge the anonymous referee for her/his valuable comments that improved this paper.


\begin{thebibliography}{49}
\expandafter\ifx\csname natexlab\endcsname\relax\def\natexlab#1{#1}\fi

\bibitem[{{Abbo} {et~al.}(2016){Abbo}, {Ofman}, {Antiochos}, {Hansteen},
  {Harra}, {Ko}, {Lapenta}, {Li}, {Riley}, {Strachan}, {von Steiger}, \&
  {Wang}}]{abb16p55}
{Abbo}, L., {Ofman}, L., {Antiochos}, S.~K., {et~al.} 2016, \ssr, 201, 55

\bibitem[{{Alterman} {et~al.}(2018){Alterman}, {Kasper}, {Stevens}, \&
  {Koval}}]{alt18p12}
{Alterman}, B.~L., {Kasper}, J.~C., {Stevens}, M.~L., \& {Koval}, A. 2018,
  \apj, 864, 112

\bibitem[{{Bame} {et~al.}(1975){Bame}, {Asbridge}, {Feldman}, {Gary}, \&
  {Montgomery}}]{bam75p73}
{Bame}, S.~J., {Asbridge}, J.~R., {Feldman}, W.~C., {Gary}, S.~P., \&
  {Montgomery}, M.~D. 1975, \grl, 2, 373

\bibitem[{{Bourouaine} \& {Chandran}(2013)}]{bou13p96}
{Bourouaine}, S., \& {Chandran}, B.~D.~G. 2013, \apj, 774, 96

\bibitem[{{Bruno} {et~al.}(2003){Bruno}, {Carbone}, {Sorriso-Valvo}, \&
  {Bavassano}}]{bru03p30}
{Bruno}, R., {Carbone}, V., {Sorriso-Valvo}, L., \& {Bavassano}, B. 2003,
  Journal of Geophysical Research (Space Physics), 108, 1130

\bibitem[{{Chandran} {et~al.}(2010){Chandran}, {Li}, {Rogers}, {Quataert}, \&
  {Germaschewski}}]{cha10p03}
{Chandran}, B.~D.~G., {Li}, B., {Rogers}, B.~N., {Quataert}, E., \&
  {Germaschewski}, K. 2010, \apj, 720, 503

\bibitem[{{Cranmer}(2014)}]{cra14p16}
{Cranmer}, S.~R. 2014, \apjs, 213, 16

\bibitem[{{Feldman} {et~al.}(1998){Feldman}, {Barraclough}, {Gosling},
  {McComas}, {Riley}, {Goldstein}, \& {Balogh}}]{fel98p47}
{Feldman}, W.~C., {Barraclough}, B.~L., {Gosling}, J.~T., {et~al.} 1998, \jgr,
  103, 14547

\bibitem[{{Fu} {et~al.}(2018){Fu}, {Madjarska}, {Li}, {Xia}, \&
  {Huang}}]{fuh18p84}
{Fu}, H., {Madjarska}, M.~S., {Li}, B., {Xia}, L., \& {Huang}, Z. 2018, \mnras,
  478, 1884

\bibitem[{{Gao} {et~al.}(2013){Gao}, {Lu}, {Li}, {Shan}, \& {Wang}}]{gao13p71}
{Gao}, X., {Lu}, Q., {Li}, X., {Shan}, L., \& {Wang}, S. 2013, \apj, 764, 71

\bibitem[{{Gary} {et~al.}(2001){Gary}, {Goldstein}, \& {Steinberg}}]{gar01p55}
{Gary}, S.~P., {Goldstein}, B.~E., \& {Steinberg}, J.~T. 2001, \jgr, 106, 24955

\bibitem[{{Gary} {et~al.}(2005){Gary}, {Smith}, \& {Skoug}}]{gar05p08}
{Gary}, S.~P., {Smith}, C.~W., \& {Skoug}, R.~M. 2005, Journal of Geophysical
  Research (Space Physics), 110, A07108

\bibitem[{{Gary} {et~al.}(2006){Gary}, {Yin}, \& {Winske}}]{gar06p05}
{Gary}, S.~P., {Yin}, L., \& {Winske}, D. 2006, Journal of Geophysical Research
  (Space Physics), 111, A06105

\bibitem[{{Gary} {et~al.}(2000){Gary}, {Yin}, {Winske}, \&
  {Reisenfeld}}]{gar00p20}
{Gary}, S.~P., {Yin}, L., {Winske}, D., \& {Reisenfeld}, D.~B. 2000, \jgr, 105,
20989

\bibitem[{{Hansteen} \& {Velli}(2012)}]{han12p89}
{Hansteen}, V.~H., \& {Velli}, M. 2012, \ssr, 172, 89

\bibitem[{{He} {et~al.}(2015{\natexlab{a}}){He}, {Pei}, {Wang}, {Tu}, {Marsch},
  {Zhang}, \& {Salem}}]{hej15p76}
{He}, J., {Pei}, Z., {Wang}, L., {et~al.} 2015{\natexlab{a}}, \apj, 805, 176

\bibitem[{{He} {et~al.}(2015{\natexlab{b}}){He}, {Wang}, {Tu}, {Marsch}, \&
  {Zong}}]{hej15p31}
{He}, J., {Wang}, L., {Tu}, C., {Marsch}, E., \& {Zong}, Q. 2015{\natexlab{b}},
  \apjl, 800, L31

\bibitem[{{He} {et~al.}(2009){He}, {Tu}, {Marsch}, {Guo}, {Yao}, \&
  {Tian}}]{hej09p25}
{He}, J.-S., {Tu}, C.-Y., {Marsch}, E., {et~al.} 2009, \aap, 497, 525

\bibitem[{{Hollweg} \& {Isenberg}(2002)}]{hol02p47}
{Hollweg}, J.~V., \& {Isenberg}, P.~A. 2002, Journal of Geophysical Research
  (Space Physics), 107, 1147

\bibitem[{{Hoppock} {et~al.}(2018){Hoppock}, {Chandran}, {Klein}, {Mallet}, \&
  {Verscharen}}]{hop18p15}
{Hoppock}, I.~W., {Chandran}, B.~D.~G., {Klein}, K.~G., {Mallet}, A., \&
  {Verscharen}, D. 2018, Journal of Plasma Physics, 84, 905840615

\bibitem[{{Howes} {et~al.}(2018){Howes}, {McCubbin}, \& {Klein}}]{how18p05}
{Howes}, G.~G., {McCubbin}, A.~J., \& {Klein}, K.~G. 2018, Journal of Plasma
  Physics, 84, 905840105

\bibitem[{{Hundhausen} {et~al.}(1970){Hundhausen}, {Bame}, {Asbridge}, \&
  {Sydoriak}}]{hun70p43}
{Hundhausen}, A.~J., {Bame}, S.~J., {Asbridge}, J.~R., \& {Sydoriak}, S.~J.
  1970, \jgr, 75, 4643

\bibitem[{{Isenberg} \& {Hollweg}(1983)}]{ise83p23}
{Isenberg}, P.~A., \& {Hollweg}, J.~V. 1983, \jgr, 88, 3923

\bibitem[{{Johnson} \& {Cheng}(2001)}]{joh01p21}
{Johnson}, J.~R., \& {Cheng}, C.~Z. 2001, \grl, 28, 4421

\bibitem[{{Kasper} {et~al.}(2008){Kasper}, {Lazarus}, \& {Gary}}]{kas08p03}
{Kasper}, J.~C., {Lazarus}, A.~J., \& {Gary}, S.~P. 2008, Physical Review
  Letters, 101, 261103

\bibitem[{{Kasper} {et~al.}(2006){Kasper}, {Lazarus}, {Steinberg}, {Ogilvie},
  \& {Szabo}}]{kas06p05}
{Kasper}, J.~C., {Lazarus}, A.~J., {Steinberg}, J.~T., {Ogilvie}, K.~W., \&
  {Szabo}, A. 2006, Journal of Geophysical Research (Space Physics), 111,
  A03105

\bibitem[{{Kasper} {et~al.}(2013){Kasper}, {Maruca}, {Stevens}, \&
  {Zaslavsky}}]{kas13p02}
{Kasper}, J.~C., {Maruca}, B.~A., {Stevens}, M.~L., \& {Zaslavsky}, A. 2013,
  Physical Review Letters, 110, 091102

\bibitem[{{Lepping} {et~al.}(1995){Lepping}, {Ac{\~u}na}, {Burlaga}, {Farrell},
  {Slavin}, {Schatten}, {Mariani}, {Ness}, {Neubauer}, {Whang}, {Byrnes},
  {Kennon}, {Panetta}, {Scheifele}, \& {Worley}}]{lep95p07}
{Lepping}, R.~P., {Ac{\~u}na}, M.~H., {Burlaga}, L.~F., {et~al.} 1995, \ssr,
  71, 207

\bibitem[{{Li} {et~al.}(2016){Li}, {Wang}, {Belcher}, {He}, \&
  {Richardson}}]{lih1602}
{Li}, H., {Wang}, C., {Belcher}, J.~W., {He}, J., \& {Richardson}, J.~D. 2016,
  \apjl, 824, L2

\bibitem[{{Livi} {et~al.}(1986){Livi}, {Marsch}, \& {Rosenbauer}}]{liv86p45}
{Livi}, S., {Marsch}, E., \& {Rosenbauer}, H. 1986, \jgr, 91, 8045

\bibitem[{{Lu} {et~al.}(2006){Lu}, {Xia}, \& {Wang}}]{luq06p01}
{Lu}, Q.~M., {Xia}, L.~D., \& {Wang}, S. 2006, Journal of Geophysical Research
  (Space Physics), 111, A09101

\bibitem[{{Marsch} {et~al.}(1982{\natexlab{a}}){Marsch}, {Goertz}, \&
  {Richter}}]{mar82p30}
{Marsch}, E., {Goertz}, C.~K., \& {Richter}, K. 1982{\natexlab{a}}, \jgr, 87,
  5030

\bibitem[{{Marsch} {et~al.}(1982{\natexlab{b}}){Marsch}, {Rosenbauer},
  {Schwenn}, {Muehlhaeuser}, \& {Neubauer}}]{mar82p35}
{Marsch}, E., {Rosenbauer}, H., {Schwenn}, R., {Muehlhaeuser}, K.-H., \&
  {Neubauer}, F.~M. 1982{\natexlab{b}}, \jgr, 87, 35

\bibitem[{{Marsch} {et~al.}(1982{\natexlab{c}}){Marsch}, {Schwenn},
  {Rosenbauer}, {Muehlhaeuser}, {Pilipp}, \& {Neubauer}}]{mar82p52}
{Marsch}, E., {Schwenn}, R., {Rosenbauer}, H., {et~al.} 1982{\natexlab{c}},
  \jgr, 87, 52

\bibitem[{{Martinovi{\'c}} {et~al.}(2019){Martinovi{\'c}}, {Klein}, \&
  {Bourouaine}}]{mar19p43}
{Martinovi{\'c}}, M.~M., {Klein}, K.~G., \& {Bourouaine}, S. 2019, \apj, 879,
  43

\bibitem[{{Ogilvie} {et~al.}(1995){Ogilvie}, {Chornay}, {Fritzenreiter},
  {Hunsaker}, {Keller}, {Lobell}, {Miller}, {Scudder}, {Sittler}, {Torbert},
  {Bodet}, {Needell}, {Lazarus}, {Steinberg}, {Tappan}, {Mavretic}, \&
  {Gergin}}]{ogi95p55}
{Ogilvie}, K.~W., {Chornay}, D.~J., {Fritzenreiter}, R.~J., {et~al.} 1995,
  \ssr, 71, 55

\bibitem[{{Osman} {et~al.}(2012){Osman}, {Matthaeus}, {Wan}, \&
  {Rappazzo}}]{osm12p02}
{Osman}, K.~T., {Matthaeus}, W.~H., {Wan}, M., \& {Rappazzo}, A.~F. 2012,
  Physical Review Letters, 108, 261102

\bibitem[{{Perri} {et~al.}(2012){Perri}, {Goldstein}, {Dorelli}, \&
  {Sahraoui}}]{per12p01}
{Perri}, S., {Goldstein}, M.~L., {Dorelli}, J.~C., \& {Sahraoui}, F. 2012,
  Physical Review Letters, 109, 191101

\bibitem[{{Roberts} \& {Li}(2015)}]{rob15p01}
{Roberts}, O.~W., \& {Li}, X. 2015, \apj, 802, 1

\bibitem[{{Stansby} {et~al.}(2019){Stansby}, {Perrone}, {Matteini}, {Horbury},
  \& {Salem}}]{sta19p02}
{Stansby}, D., {Perrone}, D., {Matteini}, L., {Horbury}, T.~S., \& {Salem},
  C.~S. 2019, \aap, 623, L2

\bibitem[{{Vech} {et~al.}(2017){Vech}, {Klein}, \& {Kasper}}]{vec17p11}
{Vech}, D., {Klein}, K.~G., \& {Kasper}, J.~C. 2017, \apjl, 850, L11

\bibitem[{{Vech} {et~al.}(2018){Vech}, {Klein}, \& {Kasper}}]{vec18p04}
---. 2018, \apjl, 863, L4

\bibitem[{{Wang} {et~al.}(2006){Wang}, {Wu}, \& {Yoon}}]{wan06p01}
{Wang}, C.~B., {Wu}, C.~S., \& {Yoon}, P.~H. 2006, Physical Review Letters, 96,
  125001

\bibitem[{{Wang} {et~al.}(2019){Wang}, {Alexandrova}, {Perrone}, {Dunlop},
  {Dong}, {Bingham}, {Khotyaintsev}, {Russell}, {Giles}, {Torbert}, {Ergun}, \&
  {Burch}}]{wan19p22}
{Wang}, T., {Alexandrova}, O., {Perrone}, D., {et~al.} 2019, \apjl, 871, L22

\bibitem[{{Wu} \& {Yoon}(2007)}]{wuc07p01}
{Wu}, C.~S., \& {Yoon}, P.~H. 2007, Physical Review Letters, 99, 075001

\bibitem[{{Yang} {et~al.}(2017){Yang}, {Lee}, {Li}, {Luo}, {Kuo}, {Shi}, \&
  {Wu}}]{yan17p77}
{Yang}, L., {Lee}, L.~C., {Li}, J.~P., {et~al.} 2017, \apj, 850, 177

\bibitem[{{Yoon} {et~al.}(2009){Yoon}, {Wang}, \& {Wu}}]{yoo09p02}
{Yoon}, P.~H., {Wang}, C.~B., \& {Wu}, C.~S. 2009, Physics of Plasmas, 16,
  102102

\bibitem[{{Zhao} {et~al.}(2019{\natexlab{a}}){Zhao}, {Feng}, {Wu}, {Pi}, \&
  {Huang}}]{zha19p75}
{Zhao}, G.~Q., {Feng}, H.~Q., {Wu}, D.~J., {Pi}, G., \& {Huang}, J.
  2019{\natexlab{a}}, \apj, 871, 175

\bibitem[{{Zhao} {et~al.}(2019{\natexlab{b}}){Zhao}, {Li}, {Feng}, {Wu}, {Li},
  \& {Zhao}}]{zha19p00}
{Zhao}, G.~Q., {Li}, H., {Feng}, H.~Q., {et~al.} 2019{\natexlab{b}}, \apj, 884, 60

\end{thebibliography}

\end{document}